\documentclass[journal]{IEEEtran}

\ifCLASSINFOpdf
\else
   \usepackage[dvips]{graphicx}
\fi
\usepackage{cite}
\usepackage{url}
\usepackage{tabularx}
\usepackage{algorithmic,algorithm}
\usepackage{graphicx}
\usepackage{amsmath,bm,graphicx}
\usepackage{amsfonts}
\usepackage{multirow}
\usepackage{booktabs}
\usepackage{diagbox}
\usepackage{hyperref}
\begin{document}

\title{A Streamable Neural Audio Codec with Residual Scalar-Vector Quantization for Real-Time Communication}

\author{Xiao-Hang~Jiang,~Yang~Ai,~\IEEEmembership{Member,~IEEE},~Rui-Chen Zheng,~Zhen-Hua~Ling,~\IEEEmembership{Senior Member,~IEEE}
\thanks{This work was funded by the National Nature Science Foundation of China under Grant 62301521 and the Anhui Provincial Natural Science Foundation under Grant 2308085QF200. (Corresponding author: Yang Ai)}
\thanks{Xiao-Hang Jiang, Yang Ai, Rui-Chen Zheng and Zhen-Hua Ling are with the National Engineering Research Center of Speech and Language Information Processing, University of Science and Technology of China, Hefei, 230027, China (e-mail: jiang\_xiaohang@mail.ustc.edu.cn, yangai@ustc.edu.cn, zhengruichen@mail.ustc.edu.cn, zhling@ustc.edu.cn).}}

\maketitle
\begin{abstract}
This paper proposes StreamCodec, a streamable neural audio codec designed for real-time communication. 
StreamCodec adopts a fully causal, symmetric encoder-decoder structure and operates in the modified discrete cosine transform (MDCT) domain, aiming for low-latency inference and real-time efficient generation. 
To improve codebook utilization efficiency and compensate for the audio quality loss caused by structural causality, StreamCodec introduces a novel residual scalar-vector quantizer (RSVQ). The RSVQ sequentially connects scalar quantizers and improved vector quantizers in a residual manner, constructing coarse audio contours and refining acoustic details, respectively.
Experimental results confirm that the proposed StreamCodec achieves decoded audio quality comparable to advanced non-streamable neural audio codecs. 
Specifically, on the 16 kHz LibriTTS dataset, StreamCodec attains a ViSQOL score of 4.30 at 1.5 kbps. It has a fixed latency of only 20 ms and achieves a generation speed nearly 20 times real-time on a CPU, with a lightweight model size of just 7M parameters, making it highly suitable for real-time communication applications.

\end{abstract}

\begin{IEEEkeywords}
streamable neural audio codec, causal structure, residual scalar-vector quantizer, real-time communication
\end{IEEEkeywords}

\IEEEpeerreviewmaketitle

\begin{figure*}
    \centering
    \includegraphics[width=\linewidth]{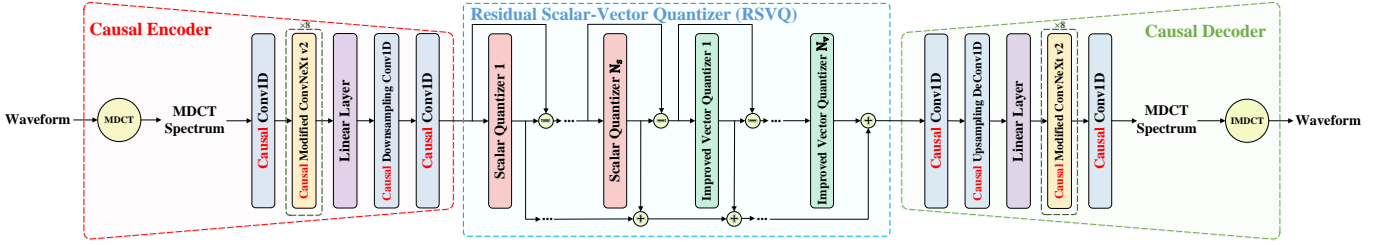}
    \caption{Overall architecture of the proposed StreamCodec.
    }
    \label{fig: overall_architecture}
\end{figure*}

\vspace{-1mm}
\section{Introduction}
\vspace{-1mm}

\IEEEPARstart{A}{udio} codec is an important signal processing technology aiming to discretize the audio signal with minimal bit usage while preserving the highest possible decoded audio quality. 
It plays an important role in various fields, such as real-time communication \cite{salami1994toll}, speech language models (SLMs) \cite{zhang2024speechtokenizer,wang2023neural,borsos2023audiolm,yang2024uniaudio}, etc.



With the development of deep learning, neural audio codecs \cite{zeghidour2021soundstream,defossez2023high,yang2023hifi,wu2023audiodec,kumar2024high,ai2024apcodec,jiang2024mdctcodec} have emerged and outperformed traditional methods such as Opus \cite{valin2013high} and EVS \cite{dietz2015overview}.
Neural audio codecs typically consist of an encoder, a decoder, and a residual vector quantizer (RVQ) \cite{vasuki2006review} equipped with trainable codebooks. 
According to the type of coding object, neural audio codecs can be divided into two categories. 
One category comprises waveform-coding-based neural audio codecs, such as SoundStream \cite{zeghidour2021soundstream}, Encodec \cite{defossez2023high}, HiFi-Codec \cite{yang2023hifi}, AudioDec \cite{wu2023audiodec} and DAC \cite{kumar2024high}, which directly discretize the audio waveform in time domain. 
These methods require hundreds of upsampling and downsampling operations, often resulting in high model complexity and low efficiency. 
The other category is based on spectral coding. 
For example, APCodec \cite{ai2024apcodec} uses a dual-path structure to encode and decode the amplitude and phase spectra, instead of the waveform, thereby improving its generation efficiency. 
Additionally, our previous work \cite{jiang2024mdctcodec} confirmed that the modified discrete cosine transform (MDCT) spectrum is more suitable for neural audio coding. Based on this finding, we proposed MDCTCodec in \cite{jiang2024mdctcodec}, which adopts a more lightweight single-path structure and a simplified loss function to achieve high-quality audio coding.

Despite the significant advancements in modern neural audio codecs, there remain several unresolved issues. 
1) Most research on neural audio codecs adopts non-causal structures, which leads to excessively high latency and limits their use in real-time communication. 
Although SoundStream \cite{zeghidour2021soundstream} and Encodec \cite{defossez2023high} support low-latency streamable inference, their coding quality and efficiency are unsatisfactory. 
AudioDec \cite{wu2023audiodec} and APCodec \cite{ai2024apcodec} also have a streamable version, but they require complicated operations, relying on HiFi-GAN vocoder \cite{kong2020hifi} and knowledge distillation strategy for assistance, respectively. 
2) The widely used RVQ \cite{vasuki2006review} strategy suffers from the codebook collapse issue \cite{guo2024addressing,zhang2024codebook}, i.e., only a few codevectors are actively utilized, wasting a large portion of the codebook resources and limiting further improvements in coding quality. 
Taking MDCTCodec \cite{jiang2024mdctcodec} as an example, the average codebook utilization rate of VQs is less than 30\%.


To overcome aforementioned challenges, this paper proposes StreamCodec, a novel streamable neural audio codec.
Built upon the MDCTCodec  \cite{jiang2024mdctcodec}, StreamCodec adopts a fully causal structure to ensure real-time processing capabilities. 
To bridge the quality gap caused by structural causality, StreamCodec introduces a novel residual scalar-vector quantizer (RSVQ). 
The RSVQ is designed based on the principle of hierarchical quantization, progressing from coarse to refined levels. It integrates the simplicity and efficiency of scalar quantization with the high precision of vector quantization, achieving a balance between computational efficiency and coding accuracy. 
Experimental results across two sampling rates and two bitrates show that our proposed StreamCodec offers high-quality audio coding with remarkable efficiency, a lightweight design, and low latency, making it an ideal solution for real-time communication applications. 
\section{Proposed Methods}
\subsection{Overview}
Figure \ref{fig: overall_architecture} provides an overview of the proposed StreamCodec. 
It comprises three main components: a causal encoder, an RSVQ and a causal decoder. 
The causal encoder first extracts the MDCT spectrum from the audio waveform, and then encodes the spectrum using a fully causal structure. 
The RSVQ structure then quantizes the encoded results. 
Finally, the causal decoder decodes the MDCT spectrum using a fully causal architecture, and the audio waveform is reconstructed through inverse MDCT (IMDCT). 
StreamCodec adopts the MDCT-based adversarial training strategy introduced in MDCTCodec \cite{jiang2024mdctcodec} while incorporating improved codebook-related training strategies, which will be introduced in Section \ref{Quantization}.

\vspace{-2mm}
\subsection{Causal Encoder and Decoder}
To enable low-latency streamable inference for real-time communication, StreamCodec employs a causal encoder and decoder which operate without relying on any future input information. 
As shown in Figure \ref{fig: overall_architecture}, the causal encoder puts two 1D causal convolutional layers respectively at the input and output ends to manipulate dimensional transformations. 
Between these layers, the encoder incorporates eight causal modified ConvNeXt v2 (MCNX2) blocks, a linear layer, and a causal 1D downsampling convolutional layer for deep feature processing. 
The causal MCNX2 blocks are adapted from the MCNX2 blocks originally introduced in APCodec \cite{ai2024apcodec} and MDCTCodec \cite{jiang2024mdctcodec} by applying causal modifications to suit real-time requirements. 
The causal decoder mirrors the structure of the encoder, with the causal 1D downsampling convolutional layer replaced by a causal 1D upsampling one.


\vspace{-2mm}
\subsection{Residual Scalar-Vector Quantizer}
\label{Quantization}

StreamCodec employs a novel RSVQ to improve coding quality. 
RSVQ consists of $N_s$ scalar quantizers (SQs) and $N_v$ improved vector quantizers (IVQs), which are connected in a residual manner. 
As described in Algorithm \ref{Algorithm}, RSVQ discretizes the output of the causal encoder $\bm{z}\in\mathbb{R}^D$ and provides the quantized result $\hat{\bm{z}}\in\mathbb{R}^D$ as the input to the causal decoder, where $D$ is the dimension of encoded/quantized features. 
The input of the first SQ is $\bm{z}$, while subsequent quantizers operate on the quantization residual of the previous quantization step. 
The final quantized result $\hat{\bm{z}}$ is obtained as the sum of the outputs of all quantizers. 


\begin{algorithm}
\caption{Residual Scalar-Vector Quantization}
\begin{algorithmic}
\label{Algorithm}
\REQUIRE{causal encoder's output $\bm{z}$, scalar quantizers $SQ_i$ ($i = 1..N_s$), improved vector quantizers $IVQ_j$ ($j = 1..N_v$)}
\ENSURE{quantized result $\hat{\bm{z}}=\bm{0.0}$, \textbf{residual} $ = \bm{z}$}
\FOR{$i = 1$ to $N_s$}
\STATE $\hat{\bm{s}}_i = SQ_i(\textbf{residual})$
\STATE $\hat{\bm{z}} = \hat{\bm{z}} + \hat{\bm{s}}_i$
\STATE $\textbf{residual} = \textbf{residual} - \hat{\bm{s}}_i$
\ENDFOR
\FOR{$j = 1$ to $N_v$}
\STATE $\hat{\bm{v}}_j = IVQ_j(\textbf{residual})$
\STATE $\hat{\bm{z}} = \hat{\bm{z}} + \hat{\bm{v}}_j$
\STATE $\textbf{residual} = \textbf{residual} - \hat{\bm{v}}_j$
\ENDFOR
\RETURN $\hat{\bm{z}}$
\end{algorithmic}
\end{algorithm}
\subsubsection{Scalar Quantizer}

In the proposed RSVQ, SQ aims to coarsely construct the audio contour. 
Inspired by \cite{mentzer2024finite,strom22_interspeech}, SQ uses rounding as the quantization principle. 
Assume that SQ discretizes an input vector $\bm{s}\in\mathbb{R}^D$ and produces a quantized result $\hat{\bm{s}}\in\mathbb{R}^D$. 
First, $\bm{s}$ undergoes a linear transformation using a trainable weight $\bm{W}_s\in\mathbb R^{B\times D}$, resulting in $\bm{s}'=\bm{W}_s\bm{s}=\left[s'_1,\dots,s'_B\right]^\top\in\mathbb R^{B}$, where $B$ is a lower dimension suitable for scalar quantization. 
The quantization process of $\bm{s}'$ is performed element-wise.
For $b$-th element $s'_b$ ($b=1,\dots,B$), the quantization result is computed as follows: 
\begin{equation}
\begin{aligned}
\hat{s}'_b &=\frac{2\cdot{round}(\tanh(s'_b + \tanh^{-1}(\frac{o_b}{h_b}))\cdot{h_b}-{o_b})}{l_b}\in\mathbb{S}_b,
\end{aligned}
\end{equation}
where
${h_b}=\frac{1.001\cdot({l_b}-1)}{2}$ and $o_b=\left| l_b \bmod 2-\frac{1}{2} \right|$. 
$l_b$ represents the number of elements in the finite scalar codebook subset $\mathbb{S}_b$, consists of $l_b$ equidistant real numbers between -1 and 1. 
Therefore, the non-trainable codebook of SQ is $\mathbb S=\mathbb S_1\times\cdots\mathbb\times\mathbb S_b\times\dots\times\mathbb S_B$ which contains $\prod_{b=1}^Bl_b$ $B$-dimensional vectors, where $\times$ denotes Cartesian product. 
The quantization result of $\bm{s}'$ is $\hat{\bm{s}}'=\left[\hat{s}'_1,\dots,\hat{s}'_B\right]^\top\in\mathbb R^{B}$. 
The SQ generates the discrete token as calculated by the following formula. 
This is a process of converting multiple non-fixed radix numbers into decimal, i.e.,
\begin{equation}
\begin{aligned}
T_s=\sum_{b = 1}^{B}\left[\left\lfloor(\hat{s}'_b+1)\cdot\frac{l_b}{2}\right\rfloor\cdot \prod_{b' = 0}^{b-1}l_{b'}\right],
\end{aligned}
\end{equation}
where $l_0=1$. 
$T_s$ can also be retrieved from $\mathbb S$ using $\hat{\bm{s}}'$.
Finally, $\hat{\bm{s}}'$ undergoes a linear transformation using a trainable weight $\bm{U}_s\in\mathbb{R}^{D\times B}$ to produce the output $\hat{\bm{s}}$ of the SQ, i.e., $\hat{\bm{s}}=\bm{U}_s\hat{\bm{s}}'$.

\begin{table*}
\centering
    \caption{Experimental results of coding quality evaluations for different neural audio codecs at two bitrates on the VCTK and LibriTTS test sets. The \textbf{bold} and \underline{underline} numbers represent the optimal and suboptimal results, respectively.}
    \resizebox{\textwidth}{!}{
    \huge
    \begin{tabular}{c | c | c c c | c c c | c c c | c c c}
\toprule
        && \multicolumn{6}{c|}{LibriTTS (16 kHz)} & \multicolumn{6}{c}{VCTK (48 kHz)}\\
        \cline{3-14}
        &Streamable& \multicolumn{3}{c|}{Target Bitrate: 1.5 kbps} & \multicolumn{3}{c|}{Target Bitrate: 2 kbps} & \multicolumn{3}{c|}{Target Bitrate: 4.5 kbps}& \multicolumn{3}{c}{Target Bitrate: 6 kbps}\\
        \cline{3-14}
         & & LSD$\downarrow$ & STOI$\uparrow$ & ViSQOL$\uparrow$ & LSD$\downarrow$ & STOI$\uparrow$ & ViSQOL$\uparrow$ & LSD$\downarrow$ & STOI$\uparrow$ & ViSQOL$\uparrow$& LSD$\downarrow$ & STOI$\uparrow$ & ViSQOL$\uparrow$ \\
\midrule
         DAC &$\times$& 0.904$\pm$0.002 & 0.915$\pm$0.002 & 3.960$\pm$0.005 & 0.897$\pm$0.001 & 0.926$\pm$0.002 & 4.064$\pm$0.005 & 0.847$\pm$0.001 & 0.874$\pm$0.002 & 3.720$\pm$0.007 & 0.839$\pm$0.001 & \textbf{0.912$\pm$0.002} & 3.797$\pm$0.007 \\
         APCodec &$\times$& 0.987$\pm$0.001 & 0.771$\pm$0.002 & 3.334$\pm$0.006 & 0.956$\pm$0.001 & 0.818$\pm$0.002 & 3.683$\pm$0.006 & \underline{0.832$\pm$0.000} & 0.864$\pm$0.002 & 3.868$\pm$0.007 & \underline{0.821$\pm$0.000} & 0.873$\pm$0.002 & 4.035$\pm$0.006 \\      
         MDCTCodec&$\times$& \underline{0.875$\pm$0.001} & \textbf{0.932$\pm$0.001} & \textbf{4.320$\pm$0.004} & \textbf{0.855$\pm$0.001} & \textbf{0.944$\pm$0.001} & \textbf{4.434$\pm$0.003} & 0.833$\pm$0.000 & \underline{0.876$\pm$0.002} & \underline{4.023$\pm$0.005} & 0.825$\pm$0.000 & 0.891$\pm$0.001 & \textbf{4.181$\pm$0.004}  \\
\midrule
         Encodec&$\checkmark$ & 0.972$\pm$0.002 & 0.888$\pm$0.002& 3.614$\pm$0.006& 0.949$\pm$0.002 & 0.912$\pm$0.002 & 3.786$\pm$0.006 & 0.906$\pm$0.001 & 0.838$\pm$0.002 & 3.415$\pm$0.010 & 0.896$\pm$0.001 & 0.846$\pm$0.002 & 3.472$\pm$0.009 \\
         AudioDec&$\checkmark$ & 0.961$\pm$0.001 & 0.709$\pm$0.002 & 3.798$\pm$0.005& 0.947$\pm$0.001 & 0.716$\pm$0.002 & 3.916$\pm$0.005 & 0.856$\pm$0.001 & 0.791$\pm$0.002 & 3.846$\pm$0.006 & 0.848$\pm$0.001 & 0.800$\pm$0.002 & 3.931$\pm$0.005 \\

         APCodec-S &$\checkmark$& 0.998$\pm$0.001 & 0.760$\pm$0.002 & 3.176$\pm$0.006 & 0.982$\pm$0.001 & 0.844$\pm$0.002 & 3.509$\pm$0.006& 0.857$\pm$0.000 & 0.844$\pm$0.002 & 3.807$\pm$0.006 & 0.844$\pm$0.000 & 0.861$\pm$0.002 & 3.89$\pm$0.006 \\
        
         MDCTCodec-S &$\checkmark$& 0.931$\pm$0.001 & 0.892$\pm$0.001 & 3.985$\pm$0.005 & 0.901$\pm$0.001 & 0.924$\pm$0.001 & 4.232$\pm$0.004& 0.838$\pm$0.000 & 0.865$\pm$0.002 & 3.900$\pm$0.006 & 0.827$\pm$0.000 & 0.891$\pm$0.002 & 4.072$\pm$0.005  \\
         StreamCodec&$\checkmark$ & \textbf{0.869$\pm$0.001} & \underline{0.926$\pm$0.001} & \underline{4.301$\pm$0.004} & \underline{0.860$\pm$0.001} & \underline{0.941$\pm$0.001} & \underline{4.393$\pm$0.004}& \textbf{0.828$\pm$0.000} & \textbf{0.887$\pm$0.002} & \textbf{4.048$\pm$0.005} & \textbf{0.820$\pm$0.000} & \underline{0.908$\pm$0.001} & \underline{4.176$\pm$0.004}  \\
    \bottomrule
    \end{tabular}}
\label{tab: Objective evaluation results}
\end{table*}

\subsubsection{Improved Vector Quantizer}
\label{Improved Vector Quantizer}

In the proposed RSVQ, IVQ aims to refine acoustic details. 
Assume that an IVQ discretizes an input vector $\bm{v}\in\mathbb{R}^D$ and produces a quantized result $\hat{\bm{v}}\in\mathbb{R}^D$. 
Similarly, $\bm{v}$ undergoes a dimensional transformation using a trainable weight $\bm{W}_v\in\mathbb R^{M\times D}$ , resulting in $\bm{v}'\in\mathbb{R}^M$, where $M$ is the dimensionality of codevectors. 
Given a trainable codebook $\mathbb V=\{\bm{v}_k|k\in\{0,1,\dots,K-1 \} \}$, where $\bm{v}_k\in\mathbb R^{M}$, and $K$ represents the number of codevectors in the codebook, the quantization result and discrete token are obtained by selecting the codevector with the closest Euclidean distance, i.e.,
\begin{equation}
\begin{aligned}
\hat{\bm{v}}',T_v=\arg\min_{(\bm{v}_k,k)} \|\bm{v}'-\bm{v}_k \|_2.
\end{aligned}
\end{equation}
Finally, $\hat{\bm{v}}'$ undergoes a dimensional recovery using a trainable weight $\bm{U}_v\in\mathbb R^{D\times M}$, resulting in the output $\hat{\bm{v}}$ of IVQ. 

Furthermore, compared to traditional VQ, IVQ introduces improvements at the training level to increase codebook utilization rate (CUR). 
Inspired by \cite{zheng2024ervq}, IVQ adopts a codebook clustering strategy during training. 
At each training step, IVQ forcibly reinitializes inactive codevectors and clusters them into the feature space to be quantized. 
Furthermore, to prevent codevectors from being activated with extremely high or low frequencies, we introduce a codebook balancing loss $\mathcal{L}_{\text{balancing}}$, defined as the cross-entropy between the posterior code distribution $\bm{P}_{post}\in\mathbb R^K$ of IVQ and the prior uniform distribution $\bm{P}_{prior}\in\mathbb R^K$. 
The posterior is approximated by the frequency with which each code is chosen during training.

\subsubsection{Bitrate Calculation}

The generated token sequence $T_s^{(1)},\dots,T_s^{(i)},\dots,T_s^{(N_s)},T_v^{(1)},\dots,T_v^{(j)},\dots,T_v^{(N_v)}$ is used for transmission in binary form in real-time communication. 
Assume that the waveform sampling rate is $f_s$ (Hz), the MDCT frame shift is $w_s$ (points) and the downsampling/upsampling rate of convolutional layers is $R$, the bitrate of StreamCodec is calculated as follows:
\begin{equation}
    \begin{aligned}
\frac{f_s}{w_s\cdot R}\cdot\left( \sum_{i=1}^{N_s}\sum_{b=1}^{B^{(i)}}  \log_2l_b^{(i)} + \sum_{j=1}^{N_v} \log_2K^{(j)}\right) (bps), 
    \end{aligned}
\end{equation}
where the superscript $i$ and $j$ represent the $i$-th SQ and the $j$-th IVQ, respectively.

\section{Experiments}
\label{sec: Experiments}

\subsection{Experimental Setting}
\label{subsec: Experimental Setting}

We conducted experiments on two datasets\footnote{Audio samples for compared neural audio codecs can be accessed at \href{https://pb20000090.github.io/StreamCodec/}{https://pb20000090.github.io/StreamCodec/}.}: LibriTTS \cite{zen2019libritts} and VCTK \cite{veaux2017superseded}.
LibriTTS comprises approximately 263 hours of audio. For our experiments, the audio was downsampled to 16 kHz (i.e., $f_s=16000$). 
We followed the official configuration \cite{zen2019libritts}, using train-clean-100 and train-clean-360 as the training set, and dev-clean and test-clean as the validation and test sets, respectively. 
VCTK contains approximately 43 hours of recordings with a sampling rate of 48 kHz (i.e., $f_s=48000$). 
The data split was consistent with that in \cite{ai2024apcodec,jiang2024mdctcodec}, taking 40936 utterances from 100 speakers as the training set and 2937 utterances from 8 unseen speakers as the test set.


StreamCodec inherited the configurations for MDCT, convolutional layers, and linear layers from MDCTCodec \cite{jiang2024mdctcodec}, including key parameters such as $w_s=40$ and $R=8$. 
For the training loss, the weight of $\mathcal{L}_{balancing}$ was set to 1. 
In the RSVQ, we adopted one SQ (i.e., $N_s=1$) and two IVQs (i.e., $N_v=2$). 
The input and output dimensions of each quantizer were both set to 32 (i.e., $D=32$). 
Experiments were conducted under two different bitrate conditions. 
For SQ, we set $l_1^{(1)}=l_2^{(1)}=l_3^{(1)}=l_4^{(1)}=l_5^{(1)}=4$ (i.e., $B^{(1)}=5$) to achieve low bitrate compression, and set $l_1^{(1)}=l_2^{(1)}=11$, $l_3^{(1)}=l_4^{(1)}=l_5^{(1)}=10$ and $l_6^{(1)}=9$ (i.e., $B^{(1)}=6$) to achieve high bitrate compression. 
For VQs, the configurations remained consistent across two bitrate conditions, with 
$M^{(1)}=M^{(2)}=32$ and $K^{(1)}=K^{(2)}=1024$.

\vspace{-2mm}
\subsection{Comparison with Baseline Neural Audio Codecs}
\label{subsec: Comparison with Baseline Codecs}

We compared the StreamCodec with several advanced neural audio codecs. 
The non-streamable codecs included DAC \cite{kumar2024high}, APCodec \cite{ai2024apcodec} and MDCTCodec \cite{jiang2024mdctcodec}, while the streamable ones included Encodec\cite{defossez2023high}, AudioDec \cite{wu2023audiodec} and APCodec-S \cite{ai2024apcodec}. 
Additionally, we produced a streamable version of MDCTCodec, named MDCTCodec-S, by simply adapting MDCTCodec to a fully causal structure.
All streamable codecs were evaluated with fixed latencies of 20 ms and 6.67 ms for sampling rates of 16 kHz and 48 kHz, respectively. 
These codecs were all reproduced based on open-source codes\footnote{\href{https://github.com/yangdongchao/AcademiCodec}{https://github.com/yangdongchao/AcademiCodec}.}\footnote{\href{https://github.com/facebookresearch/AudioDec}{https://github.com/facebookresearch/AudioDec}.}\footnote{\href{https://github.com/descriptinc/descript-audio-codec}{https://github.com/descriptinc/descript-audio-codec}.}\footnote{\href{https://github.com/yangai520/APCodec}{https://github.com/yangai520/APCodec}.}.


For coding quality evaluation, we employed log-spectral distance (LSD), short-time objective intelligibility (STOI) \cite{taal2010short} and virtual speech quality objective listener (ViSQOL) \cite{chinen2020visqol}, which assess amplitude quality, speech intelligibility and overall audio quality, respectively. 
Addtionally, we calculated the real-time factor (RTF) \cite{jiang2024mdctcodec} on both Nvidia RTX 3090 GPU and Intel E5-2680 v3 CPU for generation efficiency evaluation. Floating point operations (FLOPs) \cite{mcmahon1986livermore} and model parameters were also analyzed to assess the computational complexity and model storage efficiency, repsectively. 

As shown in Table \ref{tab: Objective evaluation results}, regardless of the dataset or bitrate, our proposed StreamCodec consistently outperformed both the non-streamable DAC and APCodec across all quality evaluation metrics, and delivered performance comparable to the non-streamable MDCTCodec.
These results demonstrate that StreamCodec achieves low-latency streamable inference while maintaining high coding quality. 
A comparison between MDCTCodec and MDCTCodec-S reveals that forcing the causality on MDCTCodec resulted in a significant performance degradation. 
However, when comparing MDCTCodec-S with StreamCodec, it becomes clear that the proposed RSVQ effectively mitigates the quality degradation introduced by the structural causality.
Efficiency and complexity evaluations were conducted on the LibriTTS dataset at a bitrate of 1.5 kbps. 
As summarized in Table \ref{tab: efficiency}, StreamCodec achieved impressive generation efficiencies of nearly 100$\times$ real-time on a GPU and 20$\times$ real-time on a CPU, while requiring only 2.51GFLOPs and 7.21M parameters. 
These results are comparable to those of MDCTCodec-S. 
Therefore, StreamCodec offers high coding quality, exceptional generation efficiency, low latency, minimal computational power consumption, and reduced storage requirements, making it an ideal choice for real-time communication applications.



\begin{table}[!t]
  \caption{Experimental results on efficiency and complexity evaluations for streamable neural audio codecs at 1.5 kbps on the LibriTTS test set. The \textbf{bold} and \underline{underline} numbers represent the optimal and suboptimal results, respectively.}
  \label{tab: efficiency}
  \centering
  \begin{tabular}{ c  c c c c }
    \toprule
    \multirow{2}{*}{}  & \multicolumn{1}{c}{RTF} & \multicolumn{1}{c}{RTF} & \multicolumn{1}{c}{FLOPs}$\downarrow$ &  \multirow{1}{*}{Parameters}$\downarrow$   \\& (GPU) $\downarrow$ & (CPU) $\downarrow$ & & \\
    \midrule
    Encodec           &0.0149&0.232&3.861G &17.60M     \\
    AudioDec               &0.0132&0.771&26.325G &24.41M       \\
    APCodec-S                   &0.0109&0.112& 4.736G &\underline{12.09M  }      \\
    MDCTCodec-S             &\textbf{0.0096}&\textbf{0.048}& \underline{2.511G} &\textbf{7.21M}    \\
    StreamCodec             &\underline{0.0101}&\underline{0.051}& \textbf{2.510G} &\textbf{7.21M}    \\
    \bottomrule
  \end{tabular}
\end{table}

\begin{figure}[!t]
    \centering
    \includegraphics[width=\linewidth]{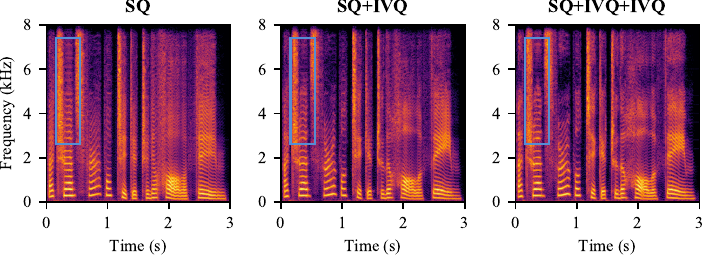}
    \caption{Comparison of audio spectrograms decoded from different quantization results by StreamCodec. SQ provides a coarse representation of the audio and subsequent IVQs refine these acoustic details. 
    }
    \label{coarse2refined}
\end{figure}

\vspace{-1.5mm}
\subsection{Analysis and Discussion on Quantization Strategies}
\label{subsec: Component Analysis}

We first analyzed the contributions of each quantizer in StreamCodec by independently feeding the quantization outputs from a single SQ, one SQ combined with the first IVQ, and one SQ combined with two IVQs into the decoder to reconstruct the audio.
The resulting spectrograms of an utterance from the LibriTTS test set are visualized in Figure \ref{coarse2refined}. 
It is evident that the single SQ provides a coarse representation of the audio, capturing the overall contour but with blurred details.
For example, the high-frequency energy distribution for the same phoneme appears roughly uniform. The subsequent IVQs refine these acoustic details, introducing distinct patterns in different frequency bins of the spectrogram.

Next, we explored the impact of various coarse and refined quantization schemes on coding performance. 
In addition to evaluating coding quality using ViSQOL, we also calculated the CUR for each quantizer and overall bitrate efficiency (BE) \cite{kumar2024high} to assess codebook utilization effectiveness. 
Experiments were conducted on the LibriTTS dataset at a bitrate of 1.5 kbps, and the results are shown in Table \ref{tab-analysis}. 
StreamCodec used one SQ for coarse quantization and two IVQs for refined quantization (i.e., index 1 in Table \ref{tab-analysis}). 
The CUR of all three quantizers in StreamCodec reached 100\%, and the BE achieved an exceptionally high value of 98.5\%.
However, when we used two SQs for refined quantization (i.e., index 2) or an IVQ for coarse quantization (i.e., index 3), both ViSQOL and BE decreased. 
This indicates that scalar quantization is more suitable for coarse quantization, while vector quantization is better suited for fine quantization. 

When we ablate the improvements for VQ (i.e., index 4), as described in Section \ref{Improved Vector Quantizer}, all metrics significantly decreased, especially the CUR of VQ, which dropped to around 50\%. 
This confirms the effectiveness of codebook clustering and the balancing loss. 
When StreamCodec used the original RVQ (i.e., index 5), all metrics significantly degraded, highlighting the advantages of the proposed RSVQ in improving both coding quality and codebook utilization performance compared to RVQ. 
Interestingly, based on the CUR of index 1 and index 4 in Table \ref{tab-analysis}, it can be observed that the decrease in the CUR of VQ also negatively affected the CUR of SQ.
However, when comparing the CUR of index 4 and index 5, it is evident that SQ helped improve the CUR of VQ. 
This indicates that scalar and vector quantization are interdependent.



\begin{table}[!t]
  \caption{Experimental results of codebook utilization performance for StreamCodec with different quantizers at 1.5 kbps on the LibriTTS test set.}
  \label{tab-analysis}
  \centering
   \resizebox{0.5\textwidth}{!}{
  \begin{tabular}{c|c|c|c|ccc|c}
\toprule
     \multirow{2}{*}{Index}&Coarse&Refined&\multirow{2}{*}{ViSQOL$\uparrow$}& \multicolumn{3}{c|}{CUR/\%($\uparrow$)}  & \multirow{2}{*}{BE/\%($\uparrow$)} \\
     &Quan.&Quan.& & \#1 & \#2 & \#3  &  \\
    \midrule
    1&SQ&IVQ+IVQ &\textbf{4.301$\pm$0.004}& 100 & 100 & 100 &  \textbf{98.5} \\
    2&SQ&SQ+SQ &4.247$\pm$0.004 & 100 &100 &100 &  97.6 \\
    3&IVQ&IVQ+IVQ& 4.090$\pm$0.005&100&100&100&96.5 \\
    4&SQ&VQ+VQ & 4.172$\pm$0.005& 43.75 & 51.95 & 57.12 &   87.8 \\
    5&VQ&VQ+VQ &3.985$\pm$0.005& 20.80 & 15.72 & 21.67 &  74.4 \\
    
    \bottomrule
    \end{tabular}
    }
\end{table}

\vspace{-1.5mm}
\section{Conclusion}
\label{sec: Conclusion}

In this paper, we propose StreamCodec, a novel streamable neural audio codec. 
StreamCodec adopts a fully causal model and introduces RSVQ to compensate for the coding quality loss caused by model causalization. 
RSVQ effectively combines the strengths of both scalar and vector quantization.
Experimental results confirm that StreamCodec is highly suitable for real-time communication, offering exceptional quality, high efficiency, low latency, and lightweight design. 
Further reducing the bitrate of StreamCodec and improving compression performance will be our future work.



\clearpage
\bibliographystyle{IEEEbib}
\bibliography{refs}

\begin{thebibliography}{10}

\bibitem{salami1994toll}
Redwan Salami, Claude Laflamme, J-P Adoul, and Dominique Massaloux,
\newblock ``A toll quality 8 kb/s speech codec for the personal communications system (pcs),''
\newblock {\em IEEE Transactions on Vehicular Technology}, vol. 43, no. 3, pp. 808--816, 1994.

\bibitem{zhang2024speechtokenizer}
Xin Zhang, Dong Zhang, Shimin Li, Yaqian Zhou, and Xipeng Qiu,
\newblock ``Speechtokenizer: {U}nified speech tokenizer for speech language models,''
\newblock in {\em Proc. ICLR}, 2024.

\bibitem{wang2023neural}
Chengyi Wang, Sanyuan Chen, Yu~Wu, Ziqiang Zhang, Long Zhou, Shujie Liu, Zhuo Chen, Yanqing Liu, Huaming Wang, Jinyu Li, et~al.,
\newblock ``Neural codec language models are zero-shot text to speech synthesizers,''
\newblock {\em arXiv preprint arXiv:2301.02111}, 2023.

\bibitem{borsos2023audiolm}
Zal{\'a}n Borsos, Rapha{\"e}l Marinier, Damien Vincent, Eugene Kharitonov, Olivier Pietquin, Matt Sharifi, Dominik Roblek, Olivier Teboul, David Grangier, Marco Tagliasacchi, et~al.,
\newblock ``Audio{LM}: a language modeling approach to audio generation,''
\newblock {\em IEEE/ACM transactions on audio, speech, and language processing}, vol. 31, pp. 2523--2533, 2023.

\bibitem{yang2024uniaudio}
Dongchao Yang, Jinchuan Tian, Xu~Tan, Rongjie Huang, Songxiang Liu, Haohan Guo, Xuankai Chang, Jiatong Shi, Jiang Bian, Zhou Zhao, et~al.,
\newblock ``Uni{A}udio: {T}owards universal audio generation with large language models,''
\newblock in {\em Proc. ICML}, 2024.

\bibitem{zeghidour2021soundstream}
Neil Zeghidour, Alejandro Luebs, Ahmed Omran, Jan Skoglund, and Marco Tagliasacchi,
\newblock ``Sound{S}tream: An end-to-end neural audio codec,''
\newblock {\em IEEE/ACM Transactions on Audio, Speech, and Language Processing}, vol. 30, pp. 495--507, 2021.

\bibitem{defossez2023high}
Alexandre D{\'e}fossez, Jade Copet, Gabriel Synnaeve, and Yossi Adi,
\newblock ``High {F}idelity {N}eural {A}udio {C}ompression,''
\newblock {\em Transactions on Machine Learning Research}, 2023.

\bibitem{yang2023hifi}
Dongchao Yang, Songxiang Liu, Rongjie Huang, Jinchuan Tian, Chao Weng, and Yuexian Zou,
\newblock ``Hi{F}i-{C}odec: Group-residual vector quantization for high fidelity audio codec,''
\newblock {\em arXiv preprint arXiv:2305.02765}, 2023.

\bibitem{wu2023audiodec}
Yi-Chiao Wu, Israel~D Gebru, Dejan Markovi{\'c}, and Alexander Richard,
\newblock ``Audio{D}ec: An open-source streaming high-fidelity neural audio codec,''
\newblock in {\em Proc. ICASSP}, 2023, pp. 1--5.

\bibitem{kumar2024high}
Rithesh Kumar, Prem Seetharaman, Alejandro Luebs, Ishaan Kumar, and Kundan Kumar,
\newblock ``High-fidelity audio compression with improved rvqgan,''
\newblock {\em Proc. NIPS}, vol. 36, 2024.

\bibitem{ai2024apcodec}
Yang Ai, Xiao-Hang Jiang, Ye-Xin Lu, Hui-Peng Du, and Zhen-Hua Ling,
\newblock ``{APCodec: A} neural audio codec with parallel amplitude and phase spectrum encoding and decoding,''
\newblock {\em IEEE/ACM Transactions on Audio, Speech, and Language Processing}, vol. 32, pp. 3256--3269, 2024.

\bibitem{jiang2024mdctcodec}
Xiao-Hang Jiang, Yang Ai, Rui-Chen Zheng, Hui-Peng Du, Ye-Xin Lu, and Zhen-Hua Ling,
\newblock ``{MDCTC}odec: A lightweight {MDCT}-based neural audio codec towards high sampling rate and low bitrate scenarios,''
\newblock in {\em Proc. SLT}, 2024, pp. 550--557.

\bibitem{valin2013high}
Jean-Marc Valin, Gregory Maxwell, Timothy~B Terriberry, and Koen Vos,
\newblock ``High-quality, low-delay music coding in the opus codec,''
\newblock in {\em Audio Engineering Society Convention 135}, 2013.

\bibitem{dietz2015overview}
Martin Dietz, Markus Multrus, Vaclav Eksler, Vladimir Malenovsky, Erik Norvell, Harald Pobloth, Lei Miao, Zhe Wang, Lasse Laaksonen, Adriana Vasilache, et~al.,
\newblock ``Overview of the {EVS} codec architecture,''
\newblock in {\em Proc. ICASSP}, 2015, pp. 5698--5702.

\bibitem{vasuki2006review}
A~Vasuki and PT~Vanathi,
\newblock ``A review of vector quantization techniques,''
\newblock {\em IEEE Potentials}, vol. 25, no. 4, pp. 39--47, 2006.

\bibitem{kong2020hifi}
Jungil Kong, Jaehyeon Kim, and Jaekyoung Bae,
\newblock ``Hi{F}i-{GAN}: Generative adversarial networks for efficient and high fidelity speech synthesis,''
\newblock in {\em Proc. NIPS}, 2020, vol.~33, pp. 17022--17033.

\bibitem{guo2024addressing}
Haohan Guo, Fenglong Xie, Dongchao Yang, Hui Lu, Xixin Wu, and Helen Meng,
\newblock ``Addressing index collapse of large-codebook speech tokenizer with dual-decoding product-quantized variational auto-encoder,''
\newblock in {\em Proc. SLT}, 2024, pp. 558--563.

\bibitem{zhang2024codebook}
Baoquan Zhang, Huaibin Wang, Chuyao Luo, Xutao Li, Guotao Liang, Yunming Ye, Xiaochen Qi, and Yao He,
\newblock ``Codebook transfer with part-of-speech for vector-quantized image modeling,''
\newblock in {\em Proc. CVPR}, 2024, pp. 7757--7766.

\bibitem{mentzer2024finite}
Fabian Mentzer, David Minnen, Eirikur Agustsson, and Michael Tschannen,
\newblock ``Finite {S}calar {Q}uantization: V{Q}-{VAE} made simple,''
\newblock in {\em Proc. ICLR}, 2024.

\bibitem{strom22_interspeech}
Nikko Strom, Haidar Khan, and Wael Hamza,
\newblock ``Squashed weight distribution for low bit quantization of deep models,''
\newblock in {\em Proc. Interspeech}, 2022, pp. 3953--3957.

\bibitem{zheng2024ervq}
Rui-Chen Zheng, Hui-Peng Du, Xiao-Hang Jiang, Yang Ai, and Zhen-Hua Ling,
\newblock ``E{RVQ}: Enhanced residual vector quantization with intra-and-inter-codebook optimization for neural audio codecs,''
\newblock {\em arXiv preprint arXiv:2410.12359}, 2024.

\bibitem{zen2019libritts}
Heiga Zen, Viet Dang, Rob Clark, Yu~Zhang, Ron~J. Weiss, Ye~Jia, Zhifeng Chen, and Yonghui Wu,
\newblock ``Libri{TTS}: {A} corpus derived from {L}ibri{S}peech for text-to-speech,''
\newblock in {\em Proc. Interspeech}, 2019, pp. 1526--1530.

\bibitem{veaux2017superseded}
Christophe Veaux, Junichi Yamagishi, Kirsten MacDonald, et~al.,
\newblock ``Superseded-{CSTR} vctk corpus: English multi-speaker corpus for {CSTR} voice cloning toolkit,''
\newblock 2017.

\bibitem{taal2010short}
Cees~H Taal, Richard~C Hendriks, Richard Heusdens, and Jesper Jensen,
\newblock ``A short-time objective intelligibility measure for time-frequency weighted noisy speech,''
\newblock in {\em Proc. ICASSP}, 2010, pp. 4214--4217.

\bibitem{chinen2020visqol}
Michael Chinen, Felicia~SC Lim, Jan Skoglund, Nikita Gureev, Feargus O'Gorman, and Andrew Hines,
\newblock ``Vi{SQOL} v3: An open source production ready objective speech and audio metric,''
\newblock in {\em Proc. QoMEX}, 2020, pp. 1--6.

\bibitem{mcmahon1986livermore}
Frank~H McMahon,
\newblock ``The {L}ivermore {F}ortran {K}ernels: A computer test of the numerical performance range,''
\newblock Tech. {R}ep., Lawrence Livermore National Lab., CA (USA), 1986.

\end{thebibliography}

\end{document}